\documentclass[]{aa}
\usepackage{times}
\usepackage{natbib}
\usepackage[dvips]{graphicx}
\bibpunct{(}{)}{;}{a}{}{,}

\def\pdot {\dot P}

\def\ltsima{$\; \buildrel < \over \sim \;$}
\def\lsim{\lower.5ex\hbox{\ltsima}}
\def\gtsima{$\; \buildrel > \over \sim \;$}
\def\gsim{\lower.5ex\hbox{\gtsima}}

\def\sax{SAX J1324--6200}
\def\4u{4U 1323--619}

\begin{document}

\title
{On the nature of the  X-ray pulsar  \sax}

\author{S. Mereghetti\inst{1}, P. Romano\inst{2}, L. Sidoli\inst{1} }

\institute {INAF, Istituto di Astrofisica Spaziale e Fisica
Cosmica Milano, via E.\ Bassini 15, I-20133 Milano, Italy
\and
INAF, Istituto di Astrofisica Spaziale e Fisica Cosmica Palermo,
Via U.\ La Malfa 153, I-90146 Palermo, Italy
      }

\offprints{S. Mereghetti, sandro@iasf-milano.inaf.it}

\date{Received 15 January 2008 / Accepted 21 February 2008}

\authorrunning{S. Mereghetti et al.}

\titlerunning{X-ray pulsar \sax}

\abstract{We present recent observations of the X--ray pulsar
\sax\ obtained in December 2007 with the Swift satellite  yielding
a significant improvement in the source localization with respect
to previous data and a new measurement of the spin period P=172.84
s. A single object consistent in colors with a highly reddened
early type star is visible in the X--ray error box. The period is
significantly longer than that obtained in 1997, indicating that
\sax\ has been spinning down at an average rate of
$\sim6\times10^{-9}$ s s$^{-1}$. We discuss the possible nature of
the source showing that it most likely belongs to the class of low
luminosity, persistent Be/neutron star binaries.
  }

\maketitle

\section{Introduction}

Little is known on the X-ray pulsar \sax ,  serendipitously
discovered in 1997 during an observation of the bright X-ray
burster \4u\  \citep{ang98}. Its X-ray spectrum, a highly absorbed
power law with photon index $\sim$1, and its pulsations period of
171 s are typical of accreting pulsars in binary systems. However,
due to the lack of an accurate position, an optical identification
has not been obtained to date, thus leading to different possible
interpretations for the nature of this source.

The great majority of accreting X-ray pulsars are High Mass X-ray
Binary Systems (HMXRB), either with OB supergiant or Be type
companion stars. \citet{ang98} suggested that \sax\ belong to the
latter group, which is the most numerous, but a white dwarf system
of the intermediate polar class could not be completely excluded.

In February 2000 a long observation of \sax , carried out with the
ASCA satellite, led to the discovery of a secular spin down at
$\sim5\times10^{-9}$ s s$^{-1}$ \citep{lin02}. This is quite
unusual for a HMXRB, since these systems are generally spinning up
while accreting. The source flux measured with ASCA  was similar
to that seen in all the previous observations of this source,
arguing against a transient nature. \citet{lin02} also reported
marginal evidence for a periodic flux modulation at 27$\pm$1
hours, that could be interpreted as the orbital period of the
system. Based on these finding, they proposed that \sax\ is a Low
Mass X-ray Binary (LMXRB), similar to other pulsars, like GX 1+4
and 4U 1626--67, which exhibited long spin-down episodes. If
confirmed, this would be particularly interesting, since only a
few LMXRB pulsars of this kind are currently known.

\section{Observations and Data Analysis}

We performed two observations of  \sax\ with the Swift satellite
on 2007 December 21 and 30 (see Table~\ref{tab:alldata} for
details), for a total on source time of 7.4\,ks. The X-ray
Telescope (XRT) data were processed with standard procedures ({\tt
xrtpipeline} v0.11.6), filtering, and screening criteria by using
FTOOLS in the {\tt Heasoft} package (v.6.4).

SAX J1324-6200 was detected in both observations with a net count
rate of $(2.9\pm0.5)\times10^{-2}$ and $(3.8\pm0.3)\times10^{-2}$
counts s$^{-1}$ in the 0.2--10\,keV energy range. The source
position, obtained  by summing all the data, is RA(J2000$)=13^{\rm
h} 24^{\rm m} 26\fs81$, Dec(J2000$)=-62^{\circ}$ $01^{\prime}
19\farcs1$, with an error of $3\farcs8$ (90\% confidence).  The
XRT coordinates lie inside the error circles derived with BeppoSAX
(1.5 arcmin radius, \citet{ang98}) and ASCA ($\sim$1 arcmin
radius, \citet{lin02}).

UVOT data were reduced with the standard software and procedures,
using tasks {\em uvotimsum} to produce the coadded images and {\em
uvotsource} to estimate the optical/UV magnitudes (U and W1
filters). There is no detection within the XRT error circle either
in individual images (6 frames in U filter and 5 in W1 filter) or
in coadded images (6113~s in U and 1989~s in W1), down to a
magnitude of U$>$21.2~mag and W1$>$20.5~mag (both consistent with
the background limit).

Given the low rate of the source, we only considered
photon-counting (PC) data and further selected XRT grades 0--12
(\citealt{Burrows2005:XRT}). No pile-up correction was necessary.
Therefore, for our spectral analysis, we extracted the source
events from a circular region with a radius of 18 pixels (1 pixel
$\sim2\farcs37$). To estimate  the background spectrum, we
extracted the events within a source-free annular region centered
on \sax\ and with radii 60 and 100 pixels. Ancillary response
files, generated with {\tt xrtmkarf},  account for different
extraction regions, vignetting,  and PSF corrections. We used the
latest spectral redistribution matrices (v010) in the Calibration
Database maintained by HEASARC.

\begin{figure}[t]%%%%%%%%%%%%%%%%%%%%%%%%%%%%%%%%%%%%%%%%%%%%% FIGURE 
        \includegraphics[angle=270,width=8.5cm]{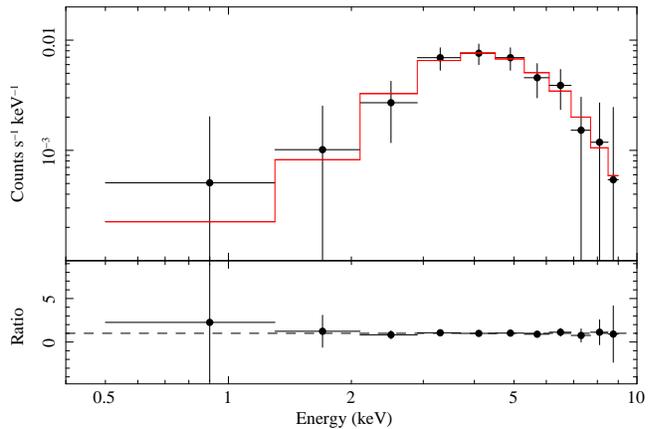}
        \caption{XRT/PC data fitted with an absorbed power law model
        (top) and data/model ratio (bottom).}
                \label{fig:spec}
    \end{figure}

We fitted the spectrum of \sax\ (216 counts)  in the 0.5--9 keV
energy range using Cash statistics and unbinned data. Adopting an
absorbed power-law model, we obtained a photon index of
$1.25\pm0.7$, a column density $N_{\rm H}=(7.5\pm3)\times 10^{22}$
cm$^{-2}$, and a Cash statistics of 579.7 using 850 PHA bins
(Fig.~\ref{fig:spec}). We calculated the goodness of the fit via
$10^4$ Montecarlo simulations, and obtained that 76.1\% of
realizations have a fit statistic $<579.7$. The observed flux is
$5.0\times10^{-12}$ erg cm$^{-2}$ s$^{-1}$ in the 1-10 keV range
(8.6$\times10^{-12}$ erg cm$^{-2}$ s$^{-1}$, corrected for the
absorption). This corresponds to an unabsorbed 1-10 keV luminosity
of 10$^{35}$d$_{10}^{2}$ erg s$^{-1}$ (where we indicate with
d$_{10}$ the distance in units of 10 kpc).

For the timing analysis, we used a source extraction radius of 25
pixels in order to increase the statistics. This resulted in 239
counts, 6\% of which we estimate are due to the background.
Considering the small number of counts we used the Z$^2$ test
\citep{buc83}, that does not require data binning,  to search for
the 171 s period. The times of arrival were corrected to the solar
system barycenter.

Swift observations consist of several snapshots (continuous
pointings at the target) whose durations, in our case,  do not
exceed $\sim1.3$\,ks. The periodicity is well visible in three of
the longest snapshots of the second observations. By analyzing
together all the data of the second observation we could derive
the best period as 172.84$\pm$0.1 s. The periodicity is only
marginally visible in the first observation, which is shorter and
yielded only $\sim$46 source counts.

\begin{figure}[t]%%%%%%%%%%%%%%%%%%%%%%%%%%%%%%%%%%%%%%%%%%%%% FIGURE 
       \includegraphics[angle=270,width=9cm]{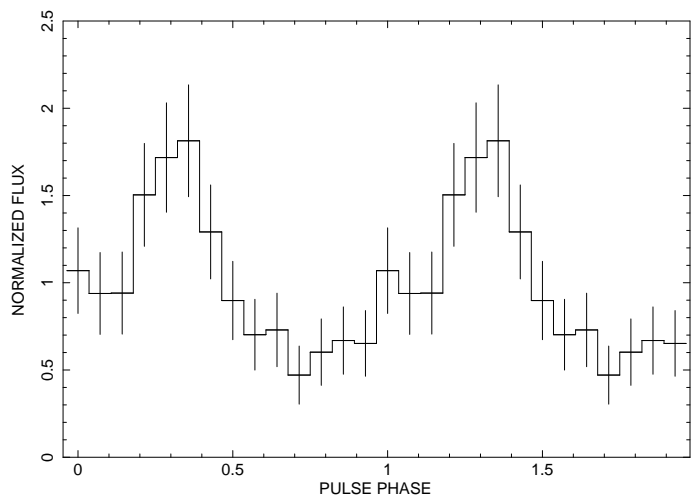}
        \caption{Folded light curve in the 0.5-9 keV energy range
        at the period P=172.84 s.}
                \label{fig:lcurve}
    \end{figure}

The folded light curve obtained by using all the data is shown in
Fig.~\ref{fig:lcurve}. The profile has a single broad peak, as
observed in the previous ASCA and BeppoSAX observations. The
pulsed fraction is about 50\%.   A hardness ratio analysis does
not show any evidence for spectral variations as a function of the
pulse phase, but this is not particularly constraining in view of
the limited number of counts.

    \begin{figure}[t]%%%%%%%%%%%%%%%%%%%%%%%%%%%%%%%%%%%%%%%%%%%%% FIGURE 
       \includegraphics[angle=270,width=9cm]{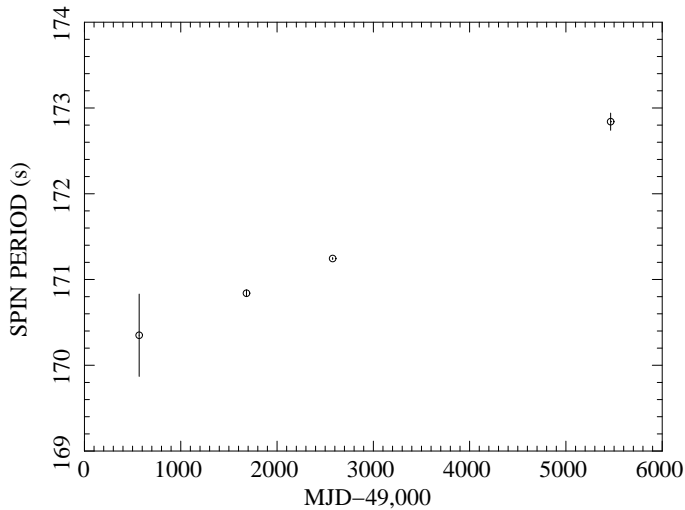}
        \caption{Long term evolution of the spin period of \sax .}
                \label{fig:pdot}
    \end{figure}

    \begin{figure}[t]%%%%%%%%%%%%%%%%%%%%%%%%%%%%%%%%%%%%%%%%%%%%% FIGURE 
       \includegraphics[angle=0,width=11.5cm]{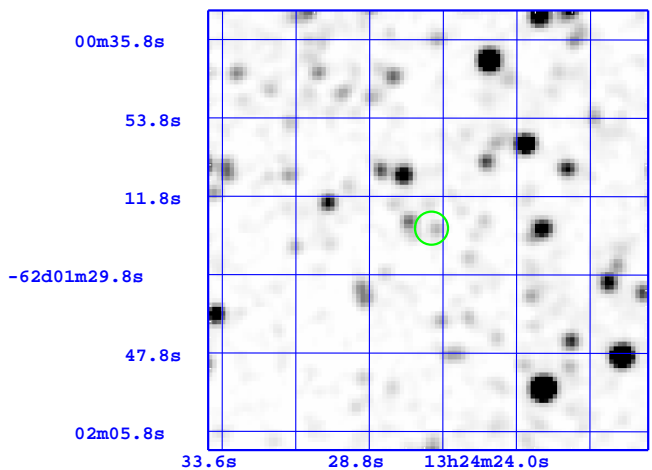}
        \caption{Image of the region of \sax\ in the $K_{\rm s}$ band. The circle is
        the XRT 90\% c.l. error region, with a radius of $3\farcs8$. North is to
        the top, East to the left. }
                \label{fig:IR}
    \end{figure}

%%%%%%%%%%%%%%%%%%% TABLES %%%%%%%%%%%%%%%%%%%%%%%%%%%%%%%%%%%%%%%%%%%

%%%%%%%%%%% A&A style
 \begin{table}
 \begin{center}
 \caption{Observation log.}
 \label{tab:alldata}
 \begin{tabular}{llll}
 \hline
 \hline
 \noalign{\smallskip}
 Sequence    & Date and Start time (UT)  & End time  & Exposure   \\
           & (yyyy-mm-dd hh:mm:ss)  & (hh:mm:ss)  &    (s)       \\
 \noalign{\smallskip}
 \hline
 \noalign{\smallskip}
37039001 &   2007-12-21 07:20:40 &    13:32:56 &   1664    \\
37039002 &   2007-12-30 00:02:31 &    08:21:57 &   5762    \\
  \noalign{\smallskip}
  \hline
  \end{tabular}
  \end{center}

  \end{table}

\section{Discussion}

All the spin period measurements of \sax\ are plotted in
Fig.~\ref{fig:pdot}.  While, as mentioned by \citet{lin02}, the
period increase between  the ASCA 2000   and the BeppoSAX values
could have been a short term fluctuation around a long term
spin-up trend, as often observed in accreting pulsars (see, e.g.,
\citet{bil97}), the new XRT measure makes this possibility less
likely. The current period P=172.84 s (at MJD=54464.2) is
consistent with the extrapolation of the previous period  and
period derivative, $\pdot$,  values. This suggests that \sax\ has
been spinning-down at a nearly constant rate of
$\sim$6$\times10^{-9}$ s s$^{-1}$ for the last ten years, and
possibly longer.

The   precise localization of \sax\ obtained with the Swift/XRT
instrument has reduced the error region area by a factor
$\sim$500, thus allowing us to exclude many of the previous
possible counterparts present in this crowded region of the
Galactic plane.

An infrared image of the field, obtained from the 2MASS All-Sky
Survey \citep{2MASS} in the $K_{\rm s}$ (2.16 $\mu$m) band, is
shown in Fig.~\ref{fig:IR}, with the XRT error circle
superimposed. Only one object is visible inside the  X-ray error
circle, at RA(J2000$)=13^{\rm h} 24^{\rm m} 26\fs65$,
Dec(J2000$)=-62^{\circ}$ $01^{\prime} 19\farcs1$.  Its $K_{\rm s}$
magnitude is $14.39\pm0.08$ mag and it is not detected in either
$J$ or $H$, implying $J-K$\gtsima2.5 and $H-K$\gtsima1.

The next closest objects are located at RA(J2000$)=13^{\rm h}
24^{\rm m} 27\fs53$, Dec(J2000$)=-62^{\circ}$ $01^{\prime}
17\farcs5$ ($J=14.15\pm 0.03$, $H=13.80\pm 0.04$, $K_{\rm
s}=13.59\pm 0.05$ mag), and RA(J2000$)=13^{\rm h} 24^{\rm m}
26\fs91$, Dec(J2000$)=-62^{\circ}$ $01^{\prime} 13\farcs1$
($J=16.66\pm 0.12$ mag, undetected in $K$ and $H$).

\sax\ lies in a very reddened region. The total optical extinction
in this direction, as provided by the NASA/IPAC Infrared Science
Archive\footnote{http://irsa.ipac.caltech.edu/applications/DUST/},
is A$_V$$\sim$29. The high column density derived from the X--ray
spectral fits also points to a rather distant and absorbed system.
The NIR colors and magnitudes of the only object detected in the
error region are consistent with highly reddened early type stars.
For example, the counterpart could be an OB supergiant with
A$_V$\gtsima15. In this case the distance should be of $\sim$30
kpc in order to be compatible with the observed magnitudes. This
would place \sax\ far at the other edge of the Galaxy, with an
X-ray luminosity of $\sim$10$^{36}$ erg s$^{-1}$.

The observed spin-down in \sax\ is not a problem in this scenario.
For example,  a decade long spin-down phase at nearly constant
X-ray flux has been observed in the supergiant system 4U~1907+09
\citep{baykal01}.

In the alternative hypothesis of a main sequence Be counterpart, a
smaller distance in the range  $\sim$5--15 kpc and A$_V$$\sim$15
would agree with the    colors and magnitudes of the candidate
counterpart. In this case   \sax\ could be located either in the
Crux galactic arm, which is tangential to this direction (Galactic
coordinates l=306.8, b=+0.61), or in the more distant Carina arm
which is crossed at $\sim$10 kpc. For the closest distances, it
would have an X--ray luminosity of a few 10$^{34}$ erg s$^{-1}$,
similar to other non-transient Be/neutron star binaries
\citep{rei99,lapalo06}, which are characterized by a relatively
small luminosity compared to transient systems in outburst. Also
among these persistent, low luminosity Be systems there are
sources which showed long periods of spin-down, similar to \sax .
For example,  X Per  has been spinning down since 1978 at an
average rate of 3.5$\times10^{-9}$ s s$^{-1}$ \citep{lapalo07}.
The absence of a strong Fe emission line\footnote{An upper limit
of 80 eV was derived on the equivalent width of a 6.4 keV line
with ASCA \citep{lin02}. Our data are consistent with this limit,
but due to the limited statistics cannot improve it. } in \sax\
fits with the properties observed in this class of persistent, low
luminosity Be systems \citep{rei99}.

Of course, both HMXRB scenarios discussed above, would argue
against the tentative\footnote{Indeed \citet{lin02} mentioned the
possibility that this periodicity be spurious, due to the unusual
shape of the light curve and the fact that the observation in
which it was detected spanned only two cycles of the putative
period} orbital period of 27 hours reported by \citep{lin02}.
Their suggestion of a spinning down LMXRB, similar to 4U1626--67
and GX1+4, is not supported by the low flux of \sax . These LMXRB
are rather luminous systems (10$^{36}$--10$^{37}$ erg s$^{-1}$)
and \sax\ should be at more than 30 kpc to have a similar
luminosity. Moreover, our refined error region excludes a bright
red giant companion star similar to the counterpart of GX1+4. In
conclusion, also in view of the HMXBs showing long-term spin-down
phases mentioned above, we believe that there is no convincing
evidence for \sax\ being a LMXRB.

\medskip

This work has been partially supported by the ASI/INAF contract
I/023/05/0.  We thank N. Gehrels and the Swift team for making
these observations possible, in particular the duty scientists and
science planners. PR thanks INAF-IASFMi, where part of the work
was carried out, for their kind hospitality.

\bibliographystyle{aa}
\bibliography{saxpap}

\begin{thebibliography}{10}
\expandafter\ifx\csname natexlab\endcsname\relax\def\natexlab#1{#1}\fi

\bibitem[{{Angelini} {et~al.}(1998){Angelini}, {Church}, {Parmar},
  {Balucinska-Church}, \& {Mineo}}]{ang98}
{Angelini}, L., {Church}, M.~J., {Parmar}, A.~N., {Balucinska-Church}, M., \&
  {Mineo}, T. 1998, \aap, 339, L41

\bibitem[{{Baykal} {et~al.}(2001){Baykal}, {Inam}, {Ali Alpar}, {in't Zand}, \&
  {Strohmayer}}]{baykal01}
{Baykal}, A., {Inam}, {\c C}., {Ali Alpar}, M., {in't Zand}, J., \&
  {Strohmayer}, T. 2001, \mnras, 327, 1269

\bibitem[{{Bildsten} {et~al.}(1997){Bildsten}, {Chakrabarty}, {Chiu}, {Finger},
  {Koh}, {Nelson}, {Prince}, {Rubin}, {Scott}, {Stollberg}, {Vaughan},
  {Wilson}, \& {Wilson}}]{bil97}
{Bildsten}, L., {Chakrabarty}, D., {Chiu}, J., {et~al.} 1997, \apjs, 113, 367

\bibitem[{{Buccheri} {et~al.}(1983){Buccheri}, {Bennett}, {Bignami}, {Bloemen},
  {Boriakoff}, {Caraveo}, {Hermsen}, {Kanbach}, {Manchester}, {Masnou},
  {Mayer-Hasselwander}, {Ozel}, {Paul}, {Sacco}, {Scarsi}, \& {Strong}}]{buc83}
{Buccheri}, R., {Bennett}, K., {Bignami}, G.~F., {et~al.} 1983, \aap, 128, 245

\bibitem[{{Burrows} {et~al.}(2005){Burrows}, {Hill}, {Nousek}, {Kennea},
  {Wells}, {Osborne}, {Abbey}, {Beardmore}, {Mukerjee}, {Short}, {Chincarini},
  {Campana}, {Citterio}, {Moretti}, {Pagani}, {Tagliaferri}, {Giommi},
  {Capalbi}, {Tamburelli}, {Angelini}, {Cusumano}, {Br{\"a}uninger}, {Burkert},
  \& {Hartner}}]{Burrows2005:XRT}
{Burrows}, D.~N., {Hill}, J.~E., {Nousek}, J.~A., {et~al.} 2005, Space Science
  Reviews, 120, 165

\bibitem[{{La Palombara} \& {Mereghetti}(2006)}]{lapalo06}
{La Palombara}, N. \& {Mereghetti}, S. 2006, \aap, 455, 283

\bibitem[{{La Palombara} \& {Mereghetti}(2007)}]{lapalo07}
---. 2007, \aap, 474, 137

\bibitem[{{Lin} {et~al.}(2002){Lin}, {Church}, {Nagase}, \&
  {Ba{\l}uci{\'n}ska-Church}}]{lin02}
{Lin}, X.~B., {Church}, M.~J., {Nagase}, F., \& {Ba{\l}uci{\'n}ska-Church}, M.
  2002, \mnras, 337, 1245

\bibitem[{{Reig} \& {Roche}(1999)}]{rei99}
{Reig}, P. \& {Roche}, P. 1999, \mnras, 306, 100

\bibitem[{{Skrutskie} {et~al.}(2006){Skrutskie}, {Cutri}, {Stiening},
  {Weinberg}, {Schneider}, {Carpenter}, {Beichman}, {Capps}, {Chester},
  {Elias}, {Huchra}, {Liebert}, {Lonsdale}, {Monet}, {Price}, {Seitzer},
  {Jarrett}, {Kirkpatrick}, {Gizis}, {Howard}, {Evans}, {Fowler}, {Fullmer},
  {Hurt}, {Light}, {Kopan}, {Marsh}, {McCallon}, {Tam}, {Van Dyk}, \&
  {Wheelock}}]{2MASS}
{Skrutskie}, M.~F., {Cutri}, R.~M., {Stiening}, R., {et~al.} 2006, \aj, 131,
  1163

\end{thebibliography}

\end{document}